# Magnetic phase diagram of HoFeO$_3$ by neutron diffraction

A. K. Ovsianikov[1,2], O.V. Usmanov[1], I. A. Zobkalo[1], V. Hutanu[2,3], S.N. Barilo[4], N.A. Liubachko[4], K.A.Shaykhutdinov[5,6], K. Yu Terentjev[5], S.V. Semenov[5,6], T. Chatterji[7], M. Meven[2,3], P. J. Brown[7], G. Roth[2], L. Peters[2], H. Deng[3], A. Wu[8].

[1] Petersburg Nuclear Physics Institute by B.P. Konstantinov of NRC «Kurchatov Institute», 188300 Gatchina, Russia.

[2] Institute of Crystallography, RWTH Aachen University, 52066 Aachen, Germany

[3] Jülich Centre for Neutron Science at Heinz Maier- Leibnitz Zentrum, Lichtenbergstrae 1, 85747 Garching, Germany

[4] Scientific-Practical Materials Research Centre NAS of Belarus, 19 P. Brovki str., Minsk, 220072, Belarus;

[5] Kirensky Institute of Physics, Federal Research Center, Krasnoyarsk 660036, Russia

[6] Siberian Federal University, Krasnoyarsk, 660071, Russia

[7] Institut Laue-Langevin, 71 Avenue des Martyrs, CS 20156 - 38042 Grenoble Cedex 9, France

[8] Institute of Ceramic, Chinese Academy of Sciences, Shanghai 200050, China

**Abstract**

Neutron diffraction studies of HoFeO$_3$ single crystals were performed under external magnetic fields. The interplay between the external magnetic fields, Dzyaloshinsky-Moria antisymmetric exchange, isotropic exchange interactions between Fe and Ho sublattices and within the Fe sublattice provides a rich magnetic phase diagram. As the result of the balance of exchange interactions inside the crystal and external magnetic fields, we found 8 different magnetic phases, induced or suppressed dependent on the external field.

**Introduction**

Rare-earth orthoferrites $R$FeO$_3$ have been studied since the 60s of the last century [1 – 3]. These compounds belong to the orthorhombic *Pnma* space group and have high Néel temperatures in the region $T_N$ = 620 – 740 K. Below $T_N$, the iron subsystem orders antiferromagnetically with a weak ferromagnetic component. With temperature decrease, the influence of the $R$ ion leads to spin-reorientation (SR) transitions, which take place in the case of magnetic rare-earth ions. There are no such transitions for the non-magnetic ions $R$ = Y, La, Lu [4]. The amount and features of SR transitions vary for different magnetic $R$ ions, as well as the characteristic temperatures $T_{SR}$. The temperature of spin-reorientation transitions ranges from $T_{SR}$ = 37 K for DyFeO$_3$ [5] to $T_{SR}$ = 480 K for SmFeO$_3$ [6]. For compounds with $R$ = Dy, Sm, Tm, Er, Yb, only one SR transition is reported while in TmFeO$_3$ and ErFeO$_3$ an additional intermediate mixed phase [7, 8] is observed. Several SR transitions were detected in crystals with $R$ = Ho, Tb [9, 10]. Spontaneous ordering of the rare-earth sublattice takes place below $T_{NR}$ ~ 10 K. Thus, the family of rare-earth orthoferrites is very suitable for studying the magnetic interaction in systems containing both 3$d$ and 4$f$ ions.

The interest in these compounds increased strongly after the prediction and subsequent discovery of their multiferroic properties. Symmetry analysis showed that orthoferrites may have ferroelectric polarization at temperatures below $T_{NR}$ [11]. Indeed, experiments have confirmed the presence of a dielectric polarization below the magnetic ordering temperature $T_{NR}$ = 5 – 10 K in DyFeO$_3$ and GdFeO$_3$ [5, 12]. However, lately electric polarization in DyFeO$_3$ was observed at higher temperature, around $T_{SR}$ = 50 – 60 K [13]. In HoFeO$_3$ spontaneous electric polarization emerges at ~ 210 K [14]. In other orthoferrites like SmFeO$_3$, YFeO$_3$ and LuFeO$_3$, electric polarization was reported even at room temperature [15 – 17]. This brings these compounds close to being useful for potential applications in switching elements, sensors, memory and other advanced technical devices with low energy consumption [18-20]. The mechanism leading to the emergence of ferroelectric polarization at high temperatures in $R$FeO$_3$ compounds is not yet understood. The Dzyaloshinskii-Moriya interaction (DMI) [21], which leads to weak ferromagnetism in the Fe3+ sublattice, could be regarded as one of the possible causes of polarization [22]. The distortion of the Fe-oxygen octahedra could play a decisive role here as it leads to a broken local symmetry.

Also, orthoferrites $R$FeO$_3$ show interesting anisotropic magnetocaloric phenomena. The magnetocaloric effect (MCE) describes the temperature change of magnetic materials in an adiabatic process caused by magnetic entropy change $\Delta S_M$ under external magnetic field. In TmFeO$_3$, the entropy change $\Delta S_M$ has a maximum of ~ 12 J/kg K at 17 K under applied field of 7 T along the $c$ axis; in TbFeO$_3$ $\Delta S_M$ reaches a value of ~ 25 J/kg K at ~ 12 K in an applied field of 7 T along the $a$ axis [23]. This is, most likely, due to a spin orientation transition of the Fe subsystems. In HoFeO$_3$ the entropy change $\Delta S_M$ has some extremums with values equal to 9 J/kg K at temperature $T$ = 53 K, $\Delta S_M$ = 15 J/kg K and $\Delta S_M$ = 18 J/kg K in the external field of 7 T at temperatures $T$ = 10 K and $T$ = 3 K respectively [24]. The first peak is also associated with a spin orientation transition in the Fe subsystem, while the second and third ones could be associated with some spin orientation transitions in the holmium subsystem. The richness of these observations clearly calls for a study of the (H, T) phase diagram of HoFeO$_3$ in external magnetic fields. This will give a better understanding of the processes inside Ho subsystem, which are presumably related to these entropy jumps $\Delta S_M$. In addition, the external magnetic field will lead to changes of the energy balance and influence the exchange interaction and anisotropy inside the holmium and iron subsystems.

The crystal structure of HoFeO$_3$ is described by the space group #62 IT [25], *Pnma* (or *Pbnm* in another setting). Below $T_N$ = 647 K, the Fe$^{3+}$ the sublattice shows an antiferromagnetic order with the strongest component along the $c$-axis, and a weak ferromagnetic component along the $b$ axis [1, 9]. As it was observed recently, the compound has two spin-reorientation transitions at $T_{SR1}$ = 53 K where the moments of the Fe$^{3+}$ ions rotate in plane from the $c$ axis to the $a$ axis and one at $T_{SR2}$ = 35 K where the strongest component of the magnetic moment is directed along the $b$ axis [9]. The Ho$^{3+}$ ordering takes place at a temperature of $T_{NR} \approx$ 10 K [26 – 28].

**Experimental**

The high quality single crystals of HoFeO$_3$ used for the neutron and magnetic studies have been grown by the flux method [29] and using the optical floating zone technique (FZ-4000, Crystal Systems Corporation) with the natural isotopes. Single crystal with parallelepiped-like shape and dimensions of 5×4×6 mm$^3$ was used for the neutron studies. The orientation and quality of the crystal obtained was controlled using the Laue technique, a typical X-ray Laue image of the HoFeO$_3$ crystal is presented in Fig. 1. Crystals for the macroscopic, X-Ray and Neutron measurements were taken from the same batch. X-Ray studies were performed at the Rigaku SmartLab diffractometer at the Petersburg Nuclear Physics Institute (PNPI). Measurements of the magnetic properties were carried out at the Krasnoyarsk Regional Center of Research Equipment of Federal Research Center «Krasnoyarsk Science Center SB RAS» on the vibrating sample magnetometer Quantum Design PPMS – 9T and Lakeshore VSM 8604.

The neutron diffraction experiments were performed at the Heinz Maier- Leibnitz Zentrum (MLZ) on the polarized neutron diffractometer POLI [30, 31]. Instrument POLI is a versatile two axes single crystal diffractometer, which can make measurements with polarized and non-polarized neutrons. For the present studies we used a non-polarized setup with the 8 T dedicated magnet. The sample was mounted in a cryomagnet, fixed on a sample stick with the glue from two sides and mechanically clamped between two aluminum plates to prevent sample rotation in the strong fields at low temperatures. In these measurements, the crystal *b*-axis was oriented vertically, while the *a* and *c* axes were laying in the horizontal scattering plane.

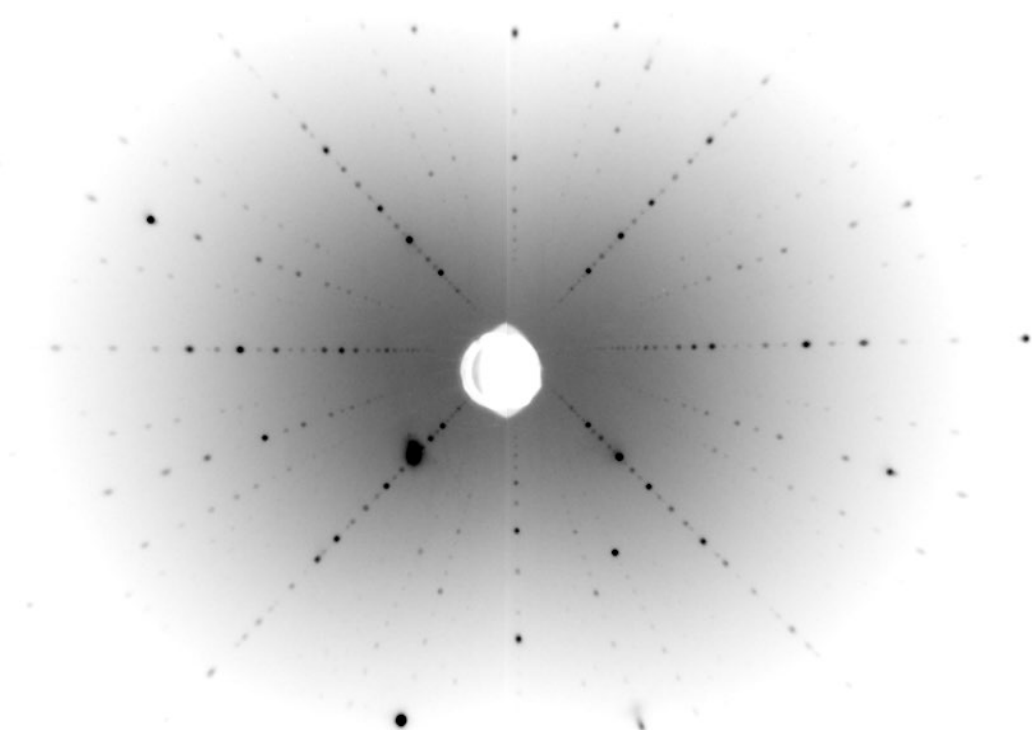

Figure 1. Typical X-ray Laue-photograph of the single crystal HoFeO$_3$ perpendicular to [010] direction, showing the high quality of the crystal.

The magnetic field was applied vertically, along the *b*-axis of the crystal. For the measurements the crystal was cooled down slowly in zero field. The measurements of the temperature dependences of 8 peaks corresponding to four different types of magnetic ordering were performed in the temperature range 2 – 68 K with 2 K steps upon heating for 7 different constant fields: 0.5, 1.25, 2.25, 3.5, 5, 6.5, 8 T. The phase transition shows only small temperature hysteresis.

**Crystal structure studies**

Powder X-Ray diffraction studies were performed at different temperatures – 300 K, 40 K and 7 K. The refinement of the obtained XRD patterns was performed for space group *Pnma* as well as for its subgroups. For this treatment, the FullProf suite [32] was used. For all temperatures parameters describing the quality of fit (like $\chi^2$, $R_f$) did not change significantly when the lower symmetry subgroups were considered. Hence, there is no evidence for a change of symmetry of the crystal structure in this temperature interval. The crystal structure (Fig. 2a) was therefore refined in space group *Pnma* at all temperatures. The most significant changes in the ion positions between 300 K and 7 K (Table 1) were observed for oxygen $O_1$. Through this anion, the super-exchange interaction along the *b* axis takes place (Fig. 2b). The exchange paths and exchange bond angles at 300 K and 7 K are presented at Table 2. It is noteworthy that both path lengths (d) and bond angles (γ) for the exchange through $O_1$ change their values to a much greater extent than those for the exchange through $O_2$, corresponding to an interaction in the *ac* plane. Interestingly, the $Fe^{3+}$ single ion anisotropy constants change their ratio from $Ab < Aac$ at 300 K to $Ab > Aac$ below 35 K [33]. This could be connected with such change of super-exchange bonds.

Table 1. Crystal structure of $HoFeO_3$ at 300 K, 7 K, *Pnma*, X-ray powder data

| 300 K | | $a = 5.5908(1)$ Å, $b = 7.6016(2)$ Å, $c = 5.2778(1)$ Å, $\chi^2 = 2.37$ | | |
|---|---|---|---|---|
| | | x | y | z |
| Fe | 4(b) | 0 | 0 | 0.5 |
| Ho | 4(c) | 0.0680(3) | 0.25 | 0.9790(4) |
| $O_1$ | 4(c) | 0.479(2) | 0.25 | 0.077(2) |
| $O_2$ | 8(d) | 0.344(2) | 0.058(1) | 0.696(2) |
| 7 K | | $a = 5.5825(1)$ Å, $b = 7.5845(1)$ Å, $c = 5.2680(1)$ Å, $\chi^2 = 4.31$ | | |
| | | x | y | z |
| Fe | 4(b) | 0 | 0 | 0.5 |
| Ho | 4(c) | 0.0693(3) | 0.25 | 0.9790(5) |
| $O_1$ | 4(c) | 0.445(2) | 0.25 | 0.097(3) |
| $O_2$ | 8(d) | 0.319(2) | 0.061(1) | 0.697(2) |

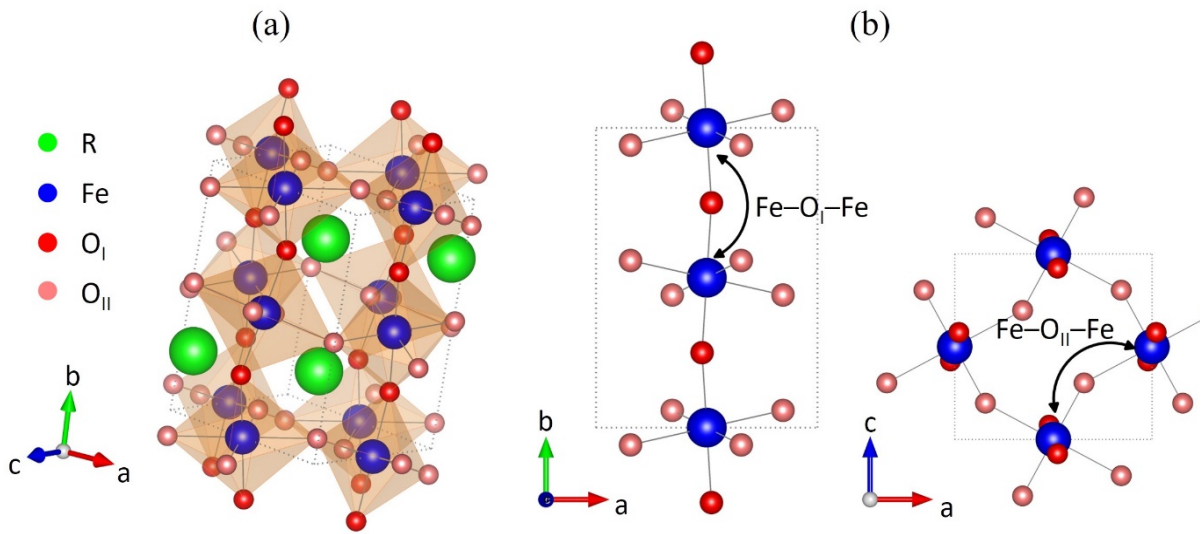

Figure 2. (a) Crystal structure $RFeO_3$; (b) Exchange paths for the nearest neighbors in the Fe sublattice. The configuration Fe–$O_I$–Fe corresponds to an exchange along the *b* axis, the configuration Fe–$O_{II}$–Fe – to an exchange in the *ac* plane.

Table 2. Exchange paths parameters at 300 K and 7 K.

|  | Fe-$O_1$-Fe | | Fe-$O_2$-Fe | |
| --- | --- | --- | --- | --- |
|  | d | γ | d | γ |
| 300 K | 3.894 (8) Å | 154.9° (7) | 4.107 (14) Å | 138.6°(5) |
| 7 K | 3.976 (10) Å | 145.1° (9) | 4.057 (16) Å | 142.2°(6) |

**Magnetization studies**

The temperature dependence of the magnetization, measured in the range 4 – 860 K, is presented in Fig. 3a. These magnetization measurements were performed on a HoFeO$_3$ single crystal at an external magnetic field $H$ = 1 kOe, parallel to the *b*-axis. The two distinct features correspond to the Néel temperature $T_N \approx$ 647 K (in good agreement with previous works [1, 3]) and to the emergence of a spin-reorientation transition in the temperature range 48 – 58 K. In order to find possible magnetic transitions in the temperature range of 2 – 100 K, the dependences of $M(T)$ at different magnetic fields were also measured (Fig. 3b). It can be seen from Fig. 3b, that a feature corresponding to spin-reorientation transition in the range 48 – 58 K is observed only at a field $H$ = 1 kOe whereas at larger fields, these dependencies become monotonous in this temperature range. The only feature of the $M(T)$ dependence noticable for the field $H$ = 30 kOe is the leveling-off at T = 4.2 K.

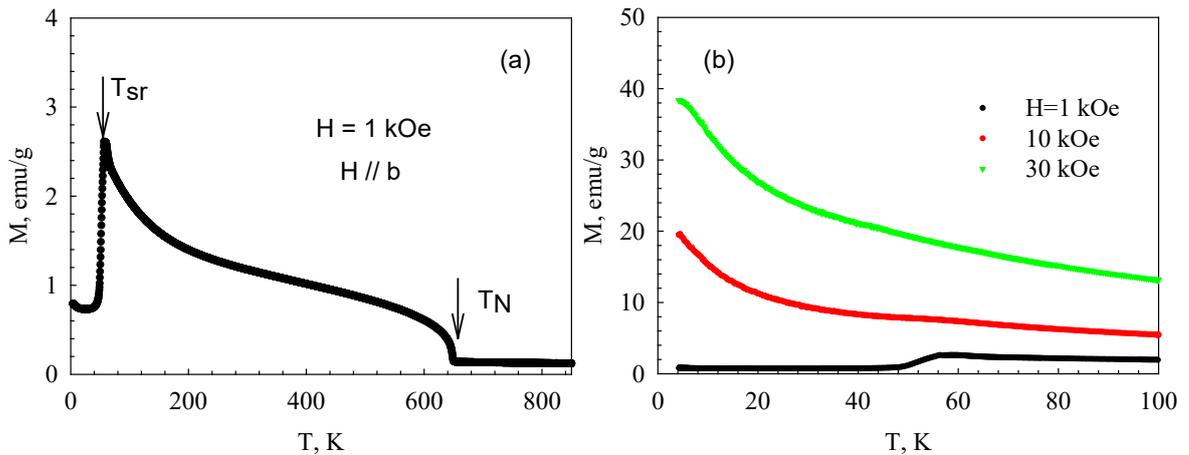

Figure 3. (a) Temperature dependence of the HoFeO$_3$ single crystal magnetization (~1kOe); (b) reduced temperature interval, M (T) for 3 different external magnetic fields.

In order to search for new features in the phase diagram of HoFeO$_3$ in the temperature range of 2 – 100 K we also measured there in detail the isotherms of the field dependences of magnetization.

Fig. 4 shows the results of measurement data of $M(H)$ in fields up to 90 kOe. It can be seen that the high-field part of the $M(H)$ dependence is linear over the entire temperature range below and above the spin-reorientation transition. The kink at small fields corresponds to the reorientation of the iron moments, and in high fields the main contribution to the magnetization comes from holmium.

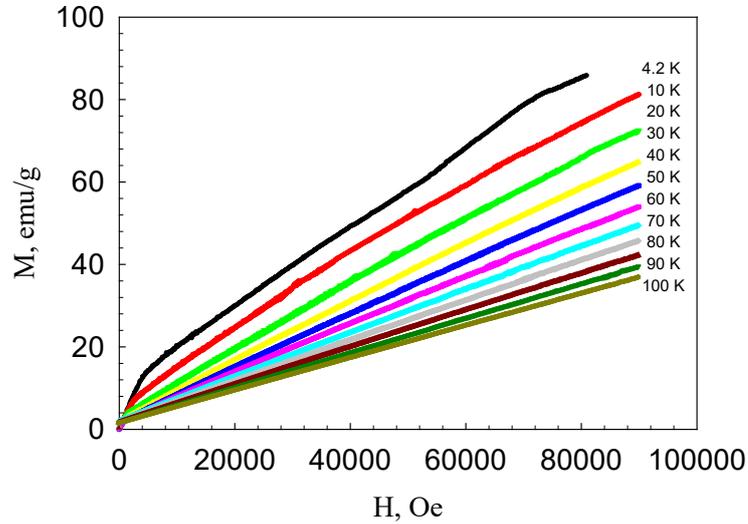

Figure 4. Field dependence of the HoFeO$_3$ single crystal magnetization at different temperatures.

The behavior of the magnetization M(H) in the region of the spin-reorientation transition at low magnetic fields was studied in detail (see Fig. 5), magnetic field $H$ was parallel to $b$ axis of the crystal. It can be seen from the Figure that, at temperatures above the spin-reorientation transition, the $M(H)$ dependences saturate rapidly, since at $T > T_{SR}$ the easy axis of the crystal is $b$ axis. At $T < T_{SR}$, already $c$ is the easy axis, and the magnetization process occurs in a wider range of fields. A small slope above the saturation of the iron moments corresponds to the influence of the paramagnetic holmium.

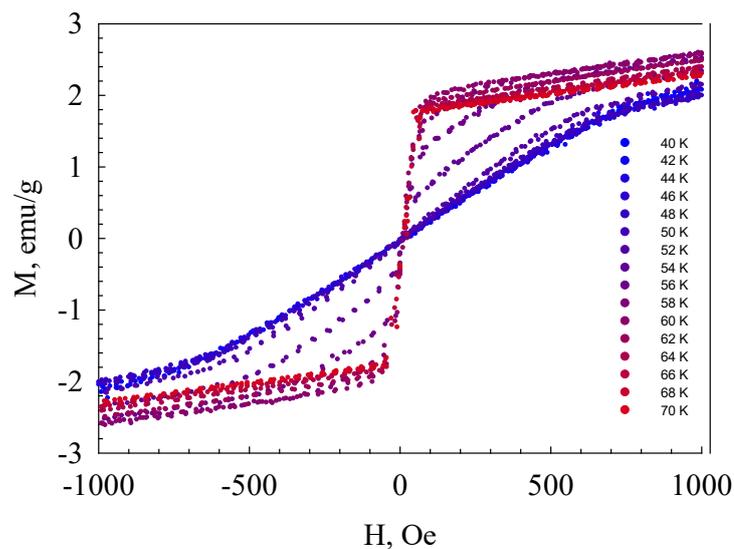

Figure 5. Field dependence of the HoFeO$_3$ single crystal magnetization in the temperature range close to $T_{SR}$.

At Figure 6 the results of the magnetization measurements along different crystallographic directions at room temperature are presented. The magnetization hysteresis with field along *b* axis connected with weak ferromagnetic component directed along *b* under the environmental condition. While the hysteresis along *c* is due to the weak magnetic component induced by the field. The spontaneous SR transition in HoFeO$_3$, which results in the reorientation of weak magnetic component from *b* to *c* axis, takes place due to the very weak Fe-Ho exchange interaction. Then weak magnetic field will be enough to induce the same reorientation transition and hysteretic behavior similar to that one for field along *b*. The difference in the shape and magnitude of the hysteresis for these two cases reflects anisotropy of the crystal. These results are in good agreement with studies of the magnetization shown in ref. [34].

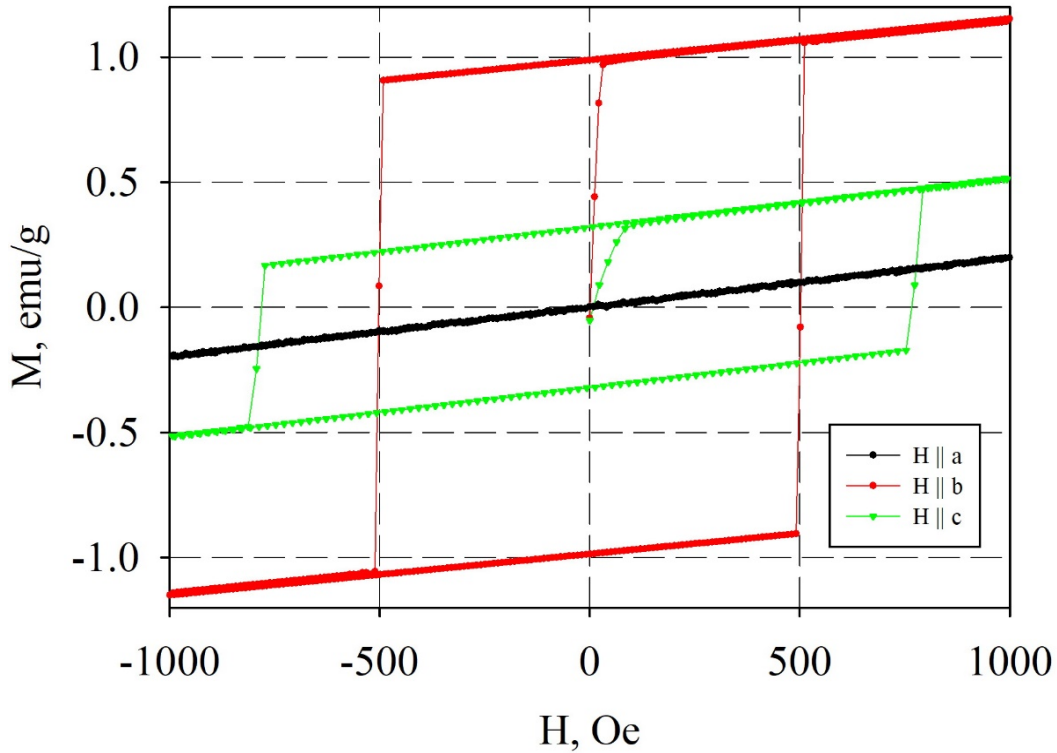

Figure 6. Field dependence of HoFeO$_3$ single crystal magnetization at room temperature.

**Magnetic symmetry analysis**

For the analysis of the magnetic scattering, the program k-SUBGROUPSMAG [35] from the Bilbao Crystallographic Server (BCS) [36 – 38] was used. The magnetic symmetry of the subgroups of space group *Pnma* with the propagation vector $\mathbf{k}$ = 0 0 0 can be described by the combination of eight one-dimensional irreducible representations (IR). Their designation in BCS notation is the following: $\Gamma_1^+, \Gamma_2^+, \Gamma_3^+, \Gamma_4^+, \Gamma_1^-, \Gamma_2^-, \Gamma_3^-, \Gamma_4^-$.

Within the Fe sublattice, the main interaction is the isotropic exchange interaction between the nearest neighbors. As the crystal cell contains four non-equivalent magnetic Fe atoms there are four types of possible collinear ordering of the Fe subsystem, which can be expressed by means of the following Bertaut notation [39]:

$$\mathbf{F} = \mathbf{S}1 + \mathbf{S}2 + \mathbf{S}3 + \mathbf{S}4,$$
$$\mathbf{G} = \mathbf{S}1 - \mathbf{S}2 + \mathbf{S}3 - \mathbf{S}4,$$
$$\mathbf{C} = \mathbf{S}1 + \mathbf{S}2 - \mathbf{S}3 - \mathbf{S}4,$$

$$A = S1 - S2 - S3 + S4,$$

where **G** describes the main antiferromagnetic component of the magnetic structure, **F** is the ferromagnetic vector and weak antiferromagnetic components **C** and **A** describe the canting of the magnetic moments. DMI leads to a canting of the sublattices and the appearance of more complex structures, which can be described using the components of collinear types of ordering. For the Fe subsystem, the decomposition of the full magnetic representation $\Gamma^{Fe}$ has the form:

$$\Gamma^{Fe} = \sum n_v^{Fe} gm_v = 3\Gamma_1^+ \oplus 3\Gamma_2^+ \oplus 3\Gamma_3^+ \oplus 3\Gamma_4^+ \tag{1}$$

For the Ho subsystem the full magnetic representation $\Gamma^{Ho}$ looks like

$$\Gamma^{Ho} = \sum n_v^{Ho} gm_v = 1\Gamma_1^+ \oplus 2\Gamma_1^- \oplus 2\Gamma_2^+ \oplus 1\Gamma_2^- \oplus 2\Gamma_3^+ \oplus 1\Gamma_3^- \oplus 1\Gamma_4^+ \oplus 2\Gamma_4^- \tag{2}$$

In the general case the magnetic moment of an atom $j$ in cell $L$ with its coordinates $\boldsymbol{R}$ and propagation vectors $\boldsymbol{k}$ may be written as a Fourier series:

$$\boldsymbol{m_j} = \sum_k \boldsymbol{S_{kj}} \cdot e^{-2\pi i k R} \tag{3}$$

The vectors $\boldsymbol{S_{kj}}$ are the Fourier components of the magnetic moment $\boldsymbol{m_j}$. They can be written as a linear combination of basics functions of irreducible representations:

$$\boldsymbol{S_{kj}} = \sum_{a,m} C_{a,m} \boldsymbol{V_{a,m}}(\boldsymbol{k}, v|j) \tag{4}$$

where the $C_{a,m}$ are the coefficients of the linear combinations, and the basic vectors $\boldsymbol{V_{a,m}}(\boldsymbol{k}, v|j)$ are constant vectors referred to the basis of the direct cell. The index $a$ varies from 1 up to the dimension of the IRs. The index $m$ varies from 1 up to the number corresponding to the number of times the representation $\Gamma_v$ is contained in the magnetic reducible representation $\Gamma$. By varying these coefficients, one can obtain all classes of magnetic structures corresponding to the symmetry of the propagation vector.

Full magnetic representation includes all irreducible magnetic representations of the subgroups of the parent group #62 for the rare-earth and iron sublattices. Table 3 shows the irreducible representations corresponding to magnetic groups and possible types of magnetic ordering for the sites Fe and Ho. Magnetic groups of lower symmetry within the orthorhombic space group can be described using the direct sum of these representations. For example, the magnetic space group *Pnm2₁* is described by the representation $\Gamma = \Gamma_1^+ \oplus \Gamma_4^-$.

Table 3. The irreducible representations and corresponding magnetic groups in the *Pnma* setting.

|  | Parent space group *Pnma* | | |
|---|---|---|---|
| Irreducible representations | Magnetic group | Site of Fe | Site of Ho |
| $\Gamma_1^+$ | *Pnma* | $G_x\ C_y\ A_z$ | $C_y$ |
| $\Gamma_2^+$ | *Pn'm'a* | $C_x\ G_y\ F_z$ | $C_x\ \ F_z$ |
| $\Gamma_3^+$ | *Pnm'a'* | $F_x\ A_y\ C_z$ | $F_x\ \ C_z$ |
| $\Gamma_4^+$ | Pn'ma' | $A_x$ $F_y$ $G_z$ | $F_y$ |
| $\Gamma_1^-$ | Pn'm'a' | | $A_x$ $G_z$ |
| $\Gamma_2^-$ | Pnma' | | $A_y$ |
| $\Gamma_3^-$ | Pn'ma | | $G_y$ |
| $\Gamma_4^-$ | Pnm'a | | $G_x$ $A_z$ |

**Magnetic scattering dependence on magnetic field**

In the $HoFeO_3$ unit cell, the $Fe^{3+}$ ions occupy the special position 4b that provides some special conditions for the contribution of different magnetic modes to Bragg reflections with corresponding Miller index parity. Thus, to reflections with $h + l$ even, $k$ odd, only mode **A** gives a contribution, to reflections with $k$ even, $h + l$ odd, only type **C** contributes, $h + l$ even, $k$ even – only type **F**, $h + l$ odd, $k$ odd, only type **G**. For the studies, two Bragg reflections for each magnetic mode were chosen: (113) and (311) for type **A**, (201) and (102) for type **C**, (002) and (200) for type **F**, (011) and (110) for type **G**. The temperature dependences of their integral intensities were measured in the external magnetic field. For all of these reflections a change in intensity with the temperature was observed. Fig. 7a shows the temperature dependence of (011) reflection intensity for all measured fields. At Fig. 7b the temperature dependences of (201) reflection for fields 0.5 T and 8 T are shown. One can see that the boundaries of the phase transitions shift down in temperature with an increase of external magnetic field. The increase of the (011) intensity at zero magnetic field in the temperature range 35 – 50 K definitely is owing to formation of phase $\Gamma_1$, which exists in that temperature range [9]. A dip in the intensity of (011) reflection in the temperature range of 50 – 55 K appears at field of 0.5 T, which is not observed at zero field (Fig. 7a). That means that the field leads to the formation of an intermediate phase during the transition below $T_{SR1}$ = 53 K. The increase of intensity still persists at fields 0 – 1.25 T and disappears with further increase in the magnetic field. No (011) intensity increase was observed below $T_{SR1}$ at the fields ≥ 2.25 T and temperature range 35-50K, where phase $\Gamma_1$ exists at zero field. Which means that phase Γ1 is suppressed by the field. The sharp increase in the intensity of the (201) reflection, which indicates an increase of the magnetic moment of holmium, shifts to higher temperature in high fields (Fig. 7b).

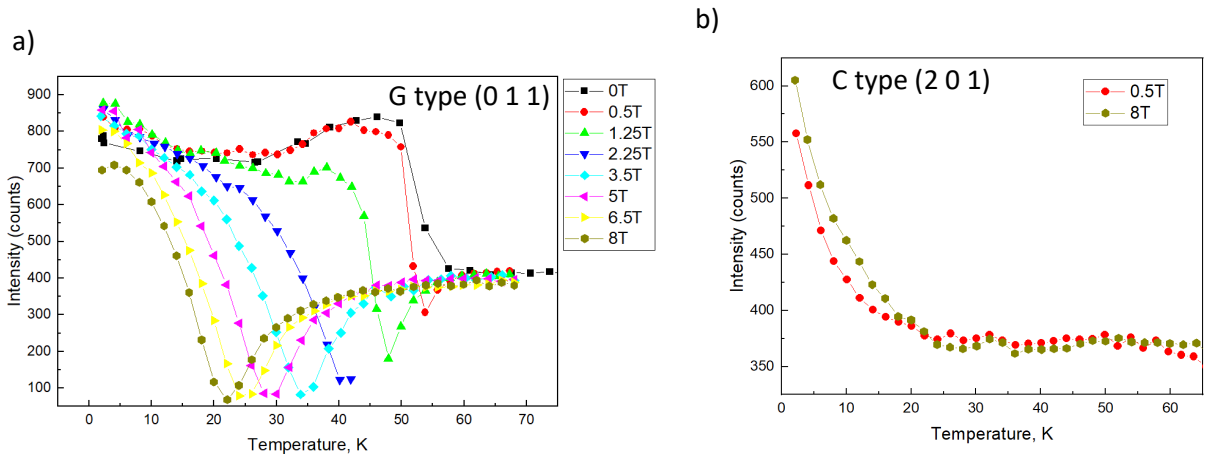

Figure 7. Temperature dependence of the integrated intensities of reflections: a) (011) of type **G** and b) (201) of type **C** in the different magnetic fields.

For magnetic structure refinement, we considered the magnetic representations shown in Table 1 and the results of refinement from reference [9]. In this way we obtained a set of magnetic phases that can be described by different magnetic representations as well as by different directions and magnitudes of the components of the magnetic moments. The resulting phase diagram is shown in Figure 8. The corresponding descriptions of the phases are given in Table 4. The magnetic moment values shown in Table 4 are rough estimates that we got when fitting 8 reflections. To refine the crystal extinction parameters, we measured these 8 peaks in zero field and used the known parameters of the magnetic moments from [9]. For all measurements in an external magnetic field, we fixed the obtained values of extinction and fitted the values of magnetic moments. We consider phases to be different also in the case they are described by the same set of irreducible presentations but with coefficients $C_{a,m}$ having different signs. Obviously, the boundaries between phases are only approximate, owing to comparatively high e.s.d.s of our magnetic structure refinements.

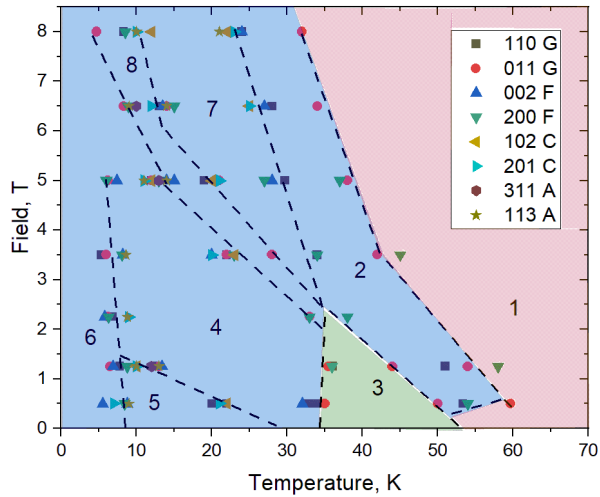

Figure 8. Magnetic phase diagram of HoFeO$_3$. Dotted line – boundary between different magnetic phases . Color area – region that are described by different magnetic representation of the Fe subsystems: red – $\Gamma_4^+$, green – $\Gamma_1^+$, blue – $\Gamma_2^+$. The numbers indicate phases as they are mentioned in the text.

Table 4. Estimates of the magnetic moments components of Fe and Ho in different magnetic phases. Magnetic moment values are given in μ$_B$.

| Magnetic representation: | | $\Gamma_4^+ \oplus \Gamma_1^-$ | $\Gamma_2^+ \oplus \Gamma_2^-$ | $\Gamma_1^+ \oplus \Gamma_1^-$ | $\Gamma_2^+ \oplus \Gamma_2^-$ | $\Gamma_2^+ \oplus \Gamma_2^-$ | $\Gamma_2^+ \oplus \Gamma_3^-$ | $\Gamma_2^+ \oplus \Gamma_2^-$ | $\Gamma_2^+ \oplus \Gamma_2^-$ |
|---|---|---|---|---|---|---|---|---|---|
| Magnetic phase | | *Pn'a'2$_1$* | *P2$_1$'2$_1$'2$_1$* | *P2$_1$2$_1$2$_1$* | *P2$_1$'2$_1$'2$_1$* | *P2$_1$'2$_1$'2$_1$* | *Pn'a2$_1$'* | *P2$_1$'2$_1$'2$_1$* | *P2$_1$'2$_1$'2$_1$* |
| Phase number | | 1 | 2 | 3 | 4 | 5 | 6 | 7 | 8 |
| Temperature | | 60 K | 40 K | 40 K | 30 K | 15 K | 5 K | 20 K | 11 K |
| Field | | 0.5 T | 5 T | 0.5 T | 0.5 T | 0.5 T | 0.5 T | 6.5 T | 6.5 T |
| Fe | Mx | 0 | 0 | 4.2(1) | 0 | 0 | 0 | 0 | 0 |
| | My | -0.7(2) | -3.3(4) | 0 | -4.5(1) | -4.8(3) | -5.2(9) | -3.3(3) | -4.3(8) |
| | Mz | -4.23(8) | -1.8 (7) | 1.2(5) | 0.2 (5) | 0.2 (5) | 0.2 (5) | -2.6(9) | -1.1(6) |

|  | Mx | 0 | 1.7(8) | -0.8(3) | 2.2(6) | 2.1(2) | 5.4(1) | 2.6(8) | 3.4(2) |
| --- | --- | --- | --- | --- | --- | --- | --- | --- | --- |
| Ho | My | 0 | 0 | 0 | 0.9(6) | 1.4(3) | 0 | 0 | 0.4(5) |
|  | Mz | 0 | 0 | -0.3(2) | 1.6(5) | -2.3(3) | 3.3(2) | -2.1(1.0) | 1.0(6) |
| R factor |  | 1.12 | 4.41 | 1.81 | 2.18 | 0.96 | 1.20 | 5.40 | 0.86 |

The best fit of phase 2 is obtained using representation $\Gamma = \Gamma_4^+$ with non-zero magnetic moment components of the iron subsystem. Nevertheless, previous studies using polarized neutrons [40] show that the *y*-component of the magnetization corresponding to the G-type reflections emerges during the transition from phase 1 to phase 3. Also, it was shown that an external magnetic field with B = 9 T at the temperature *T* = 70 K induces a magnetic moment around 1 $\mu_B$ on $Ho^{3+}$ [41]. Therefore, the magnetic space group $P2_1'2_1'2_1$ was used to describe the magnetic structure in magnetic fields. This phase is described by the representation $\Gamma = \Gamma_2^+ \oplus \Gamma_2^-$ which resolves the existence of the *y* component for G-type reflections and magnetic moments on the $Ho^{3+}$ ions. The groups *Pn'a'2$_1$* and *Pn'a2$_1$'* are more symmetric than *P2$_1$'2$_1$'2$_1$* and *P2$_1$2$_1$2$_1$*. Table 4 clearly shows that at temperatures above $T_{SR1}$ = 53 K (where only the iron subsystem is ordered) and at temperatures below $T_{NR}$ = 10 K (where the holmium has its own moment alignment) the system is in a higher symmetrical phase. Temperature range $T_{NR} < T < T_{SR1}$ is supposedly characterized by the interactions between the Fe- and Ho-subsystems and within the Ho subsystem. Along with that, the external field shifts down the temperature of the magnetic phase transitions. In addition, as can be seen from the phase diagram, an increase of the external field leads to transitions from the low-symmetry phase (№ 2 or № 3) to a highly symmetric phase (№ 1). The same situation was observed in $DyFeO_3$ where an external field leads to a transition from phase $P2_12_12_1$ to $Pn'a'2_1$ at temperatures around $T_{NR}$ [5]. However, in the case of $HoFeO_3$ this transition goes through an intermediate phase $P2_1'2_1'2_1$.

**Phase transitions in weak magnetic fields**

With an external magnetic field, the full magnetic Hamiltonian of $HoFeO_3$ will have the form:

$$H = H^{Fe-Fe} + H^{Fe-Ho} + H^{Ho-Ho} + H_{ext} \qquad (5)$$

The complex interplay, balance and competition of these interactions provide the observed sequence of phase transitions. In phase 1 (configuration Γ4), the interactions within the Ho subsystem and between the Ho and Fe subsystems can be neglected. In the absence of an external field, the $Ho^{3+}$ ions are in the exchange field of the $Fe^{3+}$ ordered subsystem, which induces an ordered magnetic moment on the holmium ions. At a temperature around $T_{SR1}$ = 53 K, the exchange interactions and easy-plane anisotropy in the *ac* plane play a main role within the iron subsystem [33]. The strongest super-exchange interaction is along the iron chains: $J_b$ = 4.9 meV, the interaction in the *ac* plane is slightly weaker: $J_{ac}$ = 4.76 meV [33]. However, the easy-plane anisotropy stabilizes the system and the moments on the iron sublattice lie in the *ac* plane with a small canting due to DMI. At $T_{SR1}$, the exchange interaction Fe-Ho increases to a level sufficient to rotate the Fe moment in the *ac* plane which yields the Γ1 structure – phase 3 in our notation.

It can be seen that already in a weak magnetic field above 0.5 T along the *b* axis, $HoFeO_3$ has an additional intermediate phase and transitions are shifted to lower temperature. It seems likely that the external field induces an additional ordered magnetic moment on the $Ho^{3+}$ ions which in turn produces an additional contribution to exchange interaction $J_{ij}^{Fe-Ho}$ between the Fe and Ho subsystems. The increase of the Fe-Ho exchange interaction produced by a field of 0.5 T provides the transition from phase 1 to phase 2. This is clearly visible in the temperature dependence of reflection (011) as a drop of its intensity, which takes place due to rotation of the Fe magnetic moment in the direction along the *b* axis. A magnetic field higher than 0.5 T influences the iron sublattice much stronger, leading to the essential changes in the alignment of the Fe moments. In this case, the iron subsystem produces a field opposite to the external one. Therefore, the temperature of transition from phase 1 to phase 2 decreases with the increase of the field. The behavior of the Ho magnetic moments reflects the competition between the external magnetic field and internal field from the Fe-sublattice. At low external field, the main influence on the Ho sublattice is exerted by the field from Fe-sublattice; thus interaction $J_{ij}^{Fe-Ho}$ rotates the iron magnetic moments from phase 1 to phase 3 through the *b* direction (see Fig. 9).

Since holmium has a highly anisotropic g-factor ($g_x$ = 6.7, $g_y$ = 1.6, $g_z$ = 3.5) [42], at fields higher than B = 2.5 T, the external field holds the Ho moments along *x* direction mainly and the anisotropic interaction $J_{ij}^{Fe-Ho}$ holds the iron magnetic moment close to the *b* axis. Thus phase 3 is suppressed by the field.

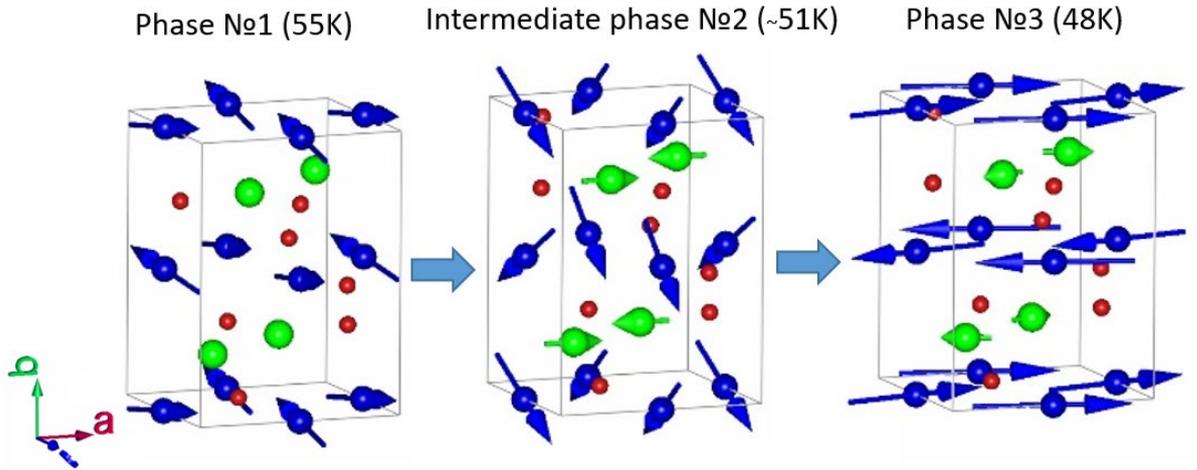

Figure 9. Scheme of magnetic structure evolution by the reorientation phase transition in $HoFeO_3$ near 50 K with magnetic field 0.5 T along [010]. Blue arrows - $Fe^{3+}$ ions, green – $Ho^{3+}$, orange spheres – $O^{2+}$.

**Low temperature phase transitions**

At the temperature $T_{SR2}$ = 35 K the increase of the induced magnetic moment of the Ho subsystem provides a subsequent increase of the exchange interaction $J_{mn}^{Ho}$ which leads to a new redistribution of the energy balance. This gives a new phase - phase 4, where the $Fe^{3+}$ moments turn closer towards the *b* direction and the system described by magnetic representation $\Gamma = \Gamma_2^+ \oplus \Gamma_2^-$ as well as for phase 2, but with a more distinct contribution from the Ho subsystem. Below $T_{SR2}$ the exchange interactions Ho-Ho and Ho-Fe become progressively more important:

$$H^{Fe-Ho} + H^{Ho-Ho} = \sum_{ij} S_i^{Fe} \cdot J_{ij}^{Fe-Ho} \cdot s_j^{Ho} - \sum_{mn} s_m^{Ho} \cdot J_{mn}^{Ho} \cdot s_n^{Ho} \quad (6)$$
$$+ \sum_{ij} D_{ij}^{Fe-Ho} \cdot (S_i^{Fe} \times s_j^{Ho}) - \sum_{mn} D_{mn}^{Ho-Ho} \cdot (s_m^{Ho} \times s_n^{Ho}) \sum_i s_i^{Ho} \cdot \boldsymbol{H}_{ext}$$

where $J_{ij}$ is the isotropic exchange, $D_{ij}^{N-M}$ the Dzyaloshinsky vectors which could consist of antisymmetric exchange and single ion anisotropy, $S_i^{Fe}$ the effective spin operator of Fe, and $s^{Ho}$ the Ho spin moment operator:

$$s^{Ho} = (g_j - 1)J \qquad (7)$$

where $g_j$ is the Lande factor, $J$ the total angular moment.

It has been established that the exchange interactions Ho-Ho and Ho-Fe have different signs [33, 43]. An increase of the ordered magnetic moment on holmium ions leads to an increase of the contribution to the energy from the exchange Ho-Ho. Most likely, the similarity of Ho-Ho and Ho-Fe interactions is, probably, the reason of the transition at $T = 24$ K (in this work, and $T = 20$ K in [9]) to phase 5, where the direction of the ferromagnetic component of Ho changes its sign (see Fig. 10a).

A magnetic field above 1 T destroys this balance, and suppresses this phase also (Fig. 10b). At temperature $T_{NR} = 10$ K $Ho^{3+}$ spontaneous magnetic ordering takes place. It leads to a new rebalancing of the energy: the exchange interaction within the rare-earth subsystem is already strong enough to hold its magnetic moment mainly along the $Ho^{3+}$ easy axis $x$ and not to follow the influence of iron sublattice and external field. Therefore, below $T_{NR} = 10$ K, the $Ho^{3+}$ moment along the hard axis $m_y \approx 0$, and the ferromagnetic component $m_z$ of $Ho^{3+}$ changes sign again (phase 6, Fig. 10a).

As it can be seen, fields above 2 T has a significant influence on the Ho-sublattice. At a field of $B = 2.5$ T, the energy of the external magnetic field is higher than the energy of Fe-Ho interaction. In this way, at temperatures above 35 K, phase 3 disappears. Simultaneously, external fields higher than 2 T are strong enough to induce $y$ component of the magnetic moment on the $Ho^{3+}$ ions. In this way a new phase 7 with strong canting of the Fe moments on that sublattice is formed,

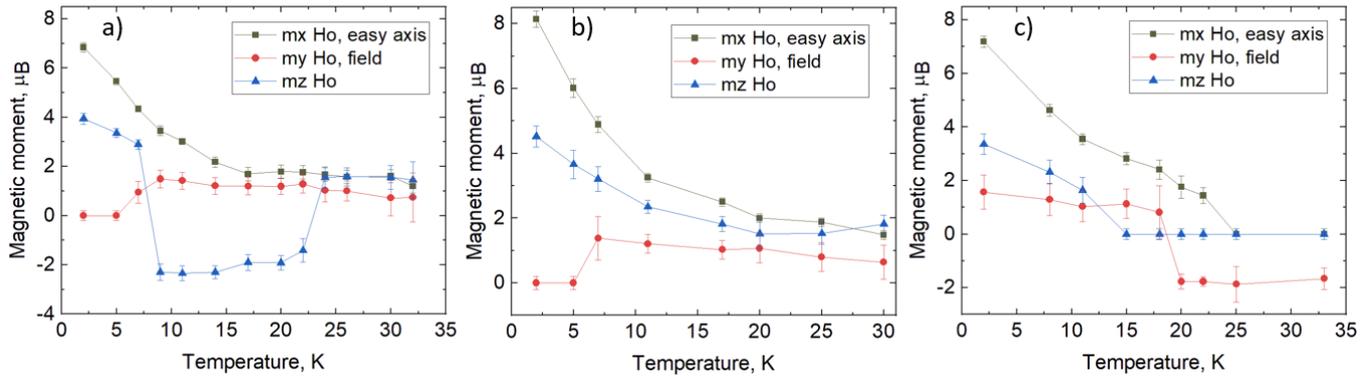

Figure 10. Temperature dependence of the $x$, $y$ and $z$ components of the Ho magnetic moments in fields of a) 0.5 T, b) 1.25 T and c) 5 T.

Phase 8 is an intermediate phase, where the Fe moments have a large component along the $b$ with a significant ferromagnetic component along $c$. A fields above 4 T lead to a canting of the Ho moments and gives a non-zero $y$-component of the Ho magnetic moment (Fig. 10c). The magnetic structures corresponding to the different magnetic phases at 20 K are shown in Figure 11.

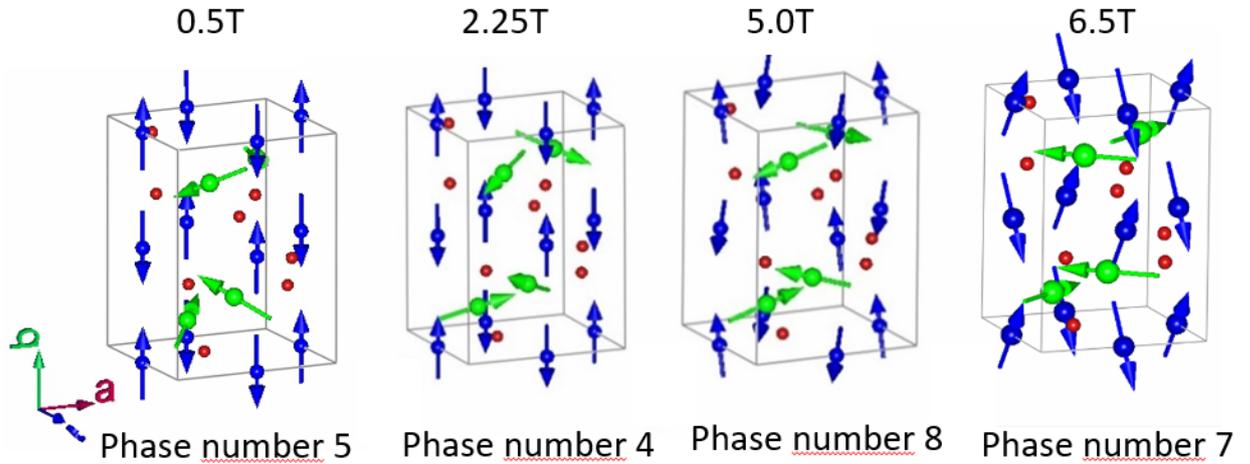

Figure 11. Schematic representation of the magnetic structures of HoFeO3 at temperature T = 20 K with increasing magnetic field (B ∥ b). Blue arrows - $Fe^{3+}$ ions, green – $Ho^{3+}$, orange spheres – $O^{2+}$.

**Conclusion**

Our studies demonstrate that HoFeO$_3$ has a rich phase diagram in an external magnetic field. The approach to the phase definition, adopted in this work is different from the conventional one, when magnetic phases described by the same set of irreducible representation considered to be the same, notwithstanding of coefficients signs. Phases defined by our approach bear different magnetic and, possibly, other different physical properties which led us to consider them to be different. The competition between the external magnetic field, the antisymmetric DMI and isotropic exchange interactions between the Fe and Ho sublattice and within the Fe sublattice leads to a complex picture of phase transitions in the rare-earth orthoferrite HoFeO$_3$. According to our considerations we can outline 8 different magnetic phases, induced or suppressed by the magnetic field. The richness of this phase diagram is the result of a very delicate balance of exchange interactions in the crystal and the external magnetic field. Such complex behavior may cause the useful functionality of rare-earth orthoferrites. The obtained results are in good agreement with the results from inelastic neutron scattering experiments [33], from which the values of exchange interactions were calculated and also with measurements of the magnetic entropy change where peaks of $\Delta S_M$ lie near the spin reorientation transition between phases 1 or 2 and 3; phases 4, 5 and 6 [24].

**Acknowledgments**

This work was supported by the Russian Foundation for Basic Research grant # 19-52-12047, and DFG grant # SA 3688/1-1. Neutron Experiments were performed at the instrument POLI jointly operated by RWTH Aachen University and Forschungszentrum Jülich at MLZ within JARA-FIT collaboration.**References**